\documentclass[iop]{emulateapj}

\usepackage{amsmath}    
\usepackage{natbib}

\usepackage{hyperref}
\usepackage{natbib}
\usepackage{color}
\usepackage{graphicx}
\usepackage{epstopdf}
\usepackage{graphicx}
\usepackage{multirow}
\newcommand{\msun}{M$_\odot$} 
 
\newcommand{\kms}{km$\cdot$s$^{-1}$}

\received{}
\accepted{}

\slugcomment{to appear in ApJ}
\shorttitle{}
\shortauthors{Wang et al.}

\begin{document}
\title{EPIC 203868608: A low-mass quadruple star system in the Upper Scorpius OB association}

\author{
Ji Wang\altaffilmark{1,2}, 
Trevor J. David\altaffilmark{1,3},
Lynne Hillenbrand\altaffilmark{1}, 
Dimitri Mawet\altaffilmark{1},
Simon Albrecht\altaffilmark{4}, and
Zibo Liu\altaffilmark{5}
} 
\email{ji.wang@caltech.edu}
\altaffiltext{1}{Department of Astronomy, California Institute of Technology, MC 249-17, 1200 E. California Blv, Pasadena, CA 91106 USA}
\altaffiltext{2}{Department of Astronomy, The Ohio State University, 100 W 18th Ave, Columbus, OH 43210 USA}
\altaffiltext{3}{Jet Propulsion Laboratory, California Institute of Technology, 4800 Oak Grove Drive, Pasadena, CA 91109, USA}
\altaffiltext{4}{Stellar Astrophysics Centre, Department of Physics and Astronomy Aarhus University, Ny Munkegade 120, 8000, Aarhus C, Denmark }
\altaffiltext{5}{Department of Astronomy \& Key Laboratory of Modern Astronomy and Astrophysics in Ministry of Education, Nanjing University,
210093, China}

\begin{abstract}

Young multiple star systems provide excellent testing grounds for theories of star formation and evolution.  EPIC 203868608 was previously studied~\citep{David2016} as a triple star system in the Upper Scorpius OB association, but the follow-up Keck NIRC2/HIRES/NIRSPAO observations reported here reveal its quadruple nature. We find that the system consists of a double-lined spectroscopic binary (SB2) Aab (M5+M5) and an eclipsing binary (EB) Bab with a total mass that is lower than that of the SB2. Furthermore, we measure the obliquity of the EB using the Doppler tomography technique during the primary eclipse. EPIC 203868608 Bab is likely on an inclined orbit with a projected obliquity of $-57^{+40}_{-36}$ degrees. The inclined orbit is used to constrain the tidal quality factor for low-mass stars and the evolution of the quadruple system. The analytic framework to infer obliquity that has been developed in this paper can be applied to other EB systems as well as transiting planets.

\end{abstract}


\section{Introduction}
\label{sec:intro}
The stellar population in the solar neighborhood is dominated by multiple star systems, mostly binaries and some triples, but with 3\% of stellar systems being quadruple and higher-order multiples~\citep{Raghavan2010}. Despite being rare, the high-order multiple systems offer unique insight into the process of star formation~\citep{Mathieu1994, Tohline2002, Reipurth2012, Duchene2013}, e.g., large scale core or filament fragmentation~\citep{Pineda2015} vs. small scale disk fragmentation~\citep{Tobin2016}. 

Among multiple stellar systems, eclipsing systems are particularly interesting because they provide an opportunity to directly measure masses and radii. The \textit{Kepler} mission~\citep{Borucki2010} greatly expanded the catalog of eclipsing binaries~\citep[EBs, ][]{Kirk2016}. In addition, higher-order multiple eclipsing stellar systems have also been discovered by the \textit{Kepler} mission~\citep{Carter2011, Derekas2011, Lehmann2012, Lehmann2016}. After the \textit{Kepler} mission, the re-purposed \textit{K2} mission~\citep{Howell2014} discovered even more high-order multiple eclipsing stellar systems~\citep[e.g.,][]{Alonso2015, Rappaport2016, Rappaport2017}. In total, the Kepler satellite greatly expands upon previously known eclipsing multiple stellar systems by revealing more than 200 high-order EBs owing to its unprecedented photometric precision and long time baseline~\citep{Conroy2014, Borkovits2016}. In the future, the Transiting Exoplanet Survey Satellite (TESS) mission~\citep{Ricker2014} and the PLATO mission~\citep{Rauer2014} will continue the trend of EB and transiting planet discoveries. 

Notably, \textit{K2} observations of the Upper Scorpius OB association have yielded a number of young eclipsing systems \citep{Kraus2015, Lodieu2015, Alonso2015}, including EPIC 203868608 \citep[][hereafter D16]{David2016}. The system previously was thought to be a hierarchical triple with a pair of eclipsing brown dwarfs. In this paper, however, we provide evidence that EPIC 203868608 is in fact a young quadruple system consisting of a double-lined spectroscopic binary (SB2) and an EB. We show that all four stars are likely to have low masses ($\lesssim$0.3~\msun) and that the EB is likely to be on an inclined orbit.

We present our observations with a suite of instruments on the Keck telescopes in \S \ref{sec:obs_red}. Results are given in \S \ref{sec:results} including stellar and orbital properties of EPIC 203868608. In \S \ref{sec:oblqt}, we focus on the obliquity of the EB in EPIC 203868608. The summary is given in \S \ref{sec:summary}. 

\section{Observation and Data Reduction}
\label{sec:obs_red}
\subsection{Keck/NIRC2}
We observed EPIC 203868608 using the Keck/NIRC2 instrument~\citep{Wizinowich2000} in laser guide star (LGS) mode. LGS was required for an acceptable adaptive optics (AO) performance because EPIC 203868608 is faint in wavefront sensing wavelengths ($r=16.3$~mag) and cannot serve as a natural guide star. We obtained AO images in the $J$ and $K_P$ bands at three epochs. The first two epochs of observations are obtained from Keck Observatory Archive (PID: N121N2L, PI: Mann, and PID: H210N2L, PI: Baranec). The first epoch was coincidentally taken during the primary eclipse (UT 2015 Jun 22, MJD 57195.42345 day). Since we set the orbital phase [0-1] to zero at the middle of the primary eclipse, the observation corresponded to an orbital phase of 0.9942. The second epoch was taken at UT 2015 Jul 25, MJD 57228.26646 day, corresponding to a orbital phase of 0.3674, when the total flux of the system is at a normal (non-eclipsing) level. We took the third epoch of observation on UT 2016 July 17 (PID: C237N2L, PI: Mawet) at an orbital phase of 0.4371, 2.94 hours from secondary eclipse. 

On UT 2015 Jun 22, two $J$-band AO frames without dithering and two $K_P$-band AO frames with a dither pitch of 0.27$^{\prime\prime}$ were taken. The total on-target time per frame was 20 seconds (1 second with 20 co-adds).

On UT 2015 Jul 25, three $J$-band AO frames were taken with a three-point dither pattern that has a throw of 2.5$^{\prime\prime}$. The lower left quadrant was avoided because it has a much higher instrumental noise than the other three quadrants on the detector. Total on-target time per frame in $J$ band was 30 seconds (1.25 seconds with 24 co-adds). Five $K_P$-band AO frames were taken with a five-point dither pattern. 
The five-point dither pattern has the target in the center of the detector and the centers of each detector quadrant. The total on-target time per frame in $K_P$ band was 60 seconds (1.25 seconds with 48 co-adds). 

On UT 2016 Jul 17, we took three $K_P$-band AO frames with a three-point dither pattern that has a throw of 2.5$^{\prime\prime}$. Total on-target time per frame was 20 seconds (2 seconds with 10 co-adds). {{All Keck NIRC2 observations and other observations (HIRES and NIRSPAO) on EPIC 203868608 are summarized in Table \ref{table:obs_summary}. }}

\begin{figure*}
  \centering
  \begin{tabular}[b]{cc}
    \includegraphics[width=.5\linewidth]{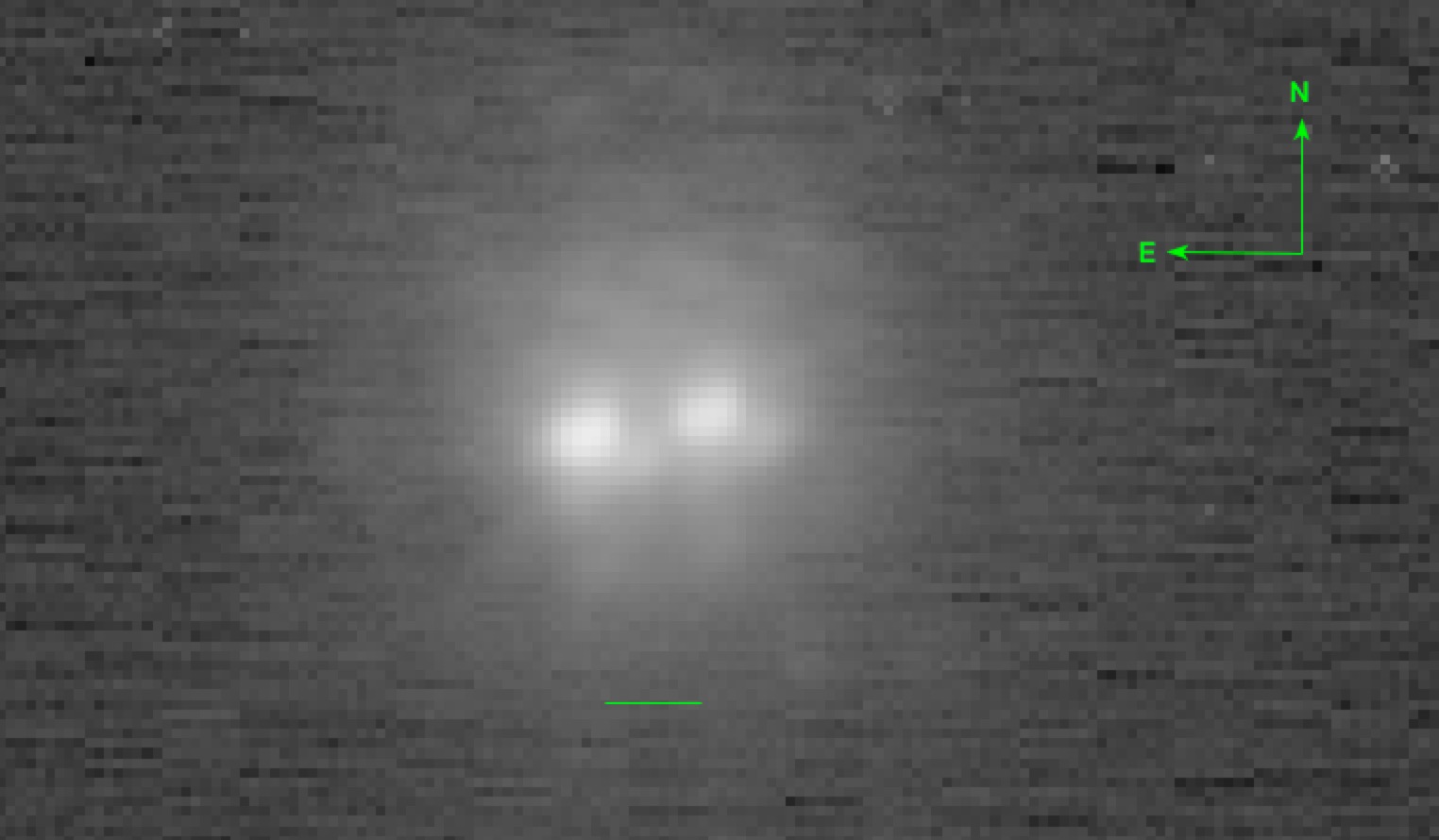} & \includegraphics[width=.48\linewidth]{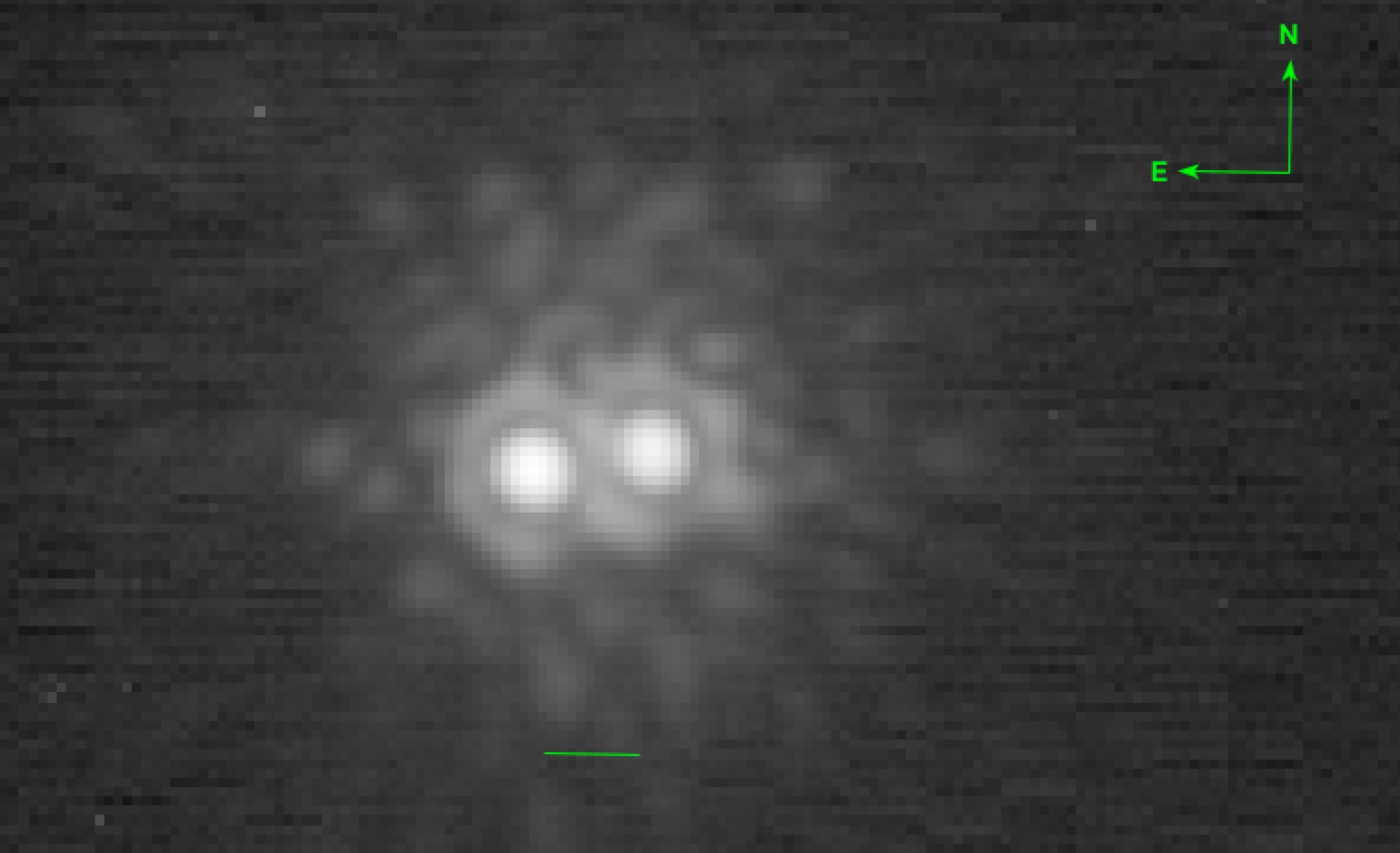} \\
    \includegraphics[width=.5\linewidth]{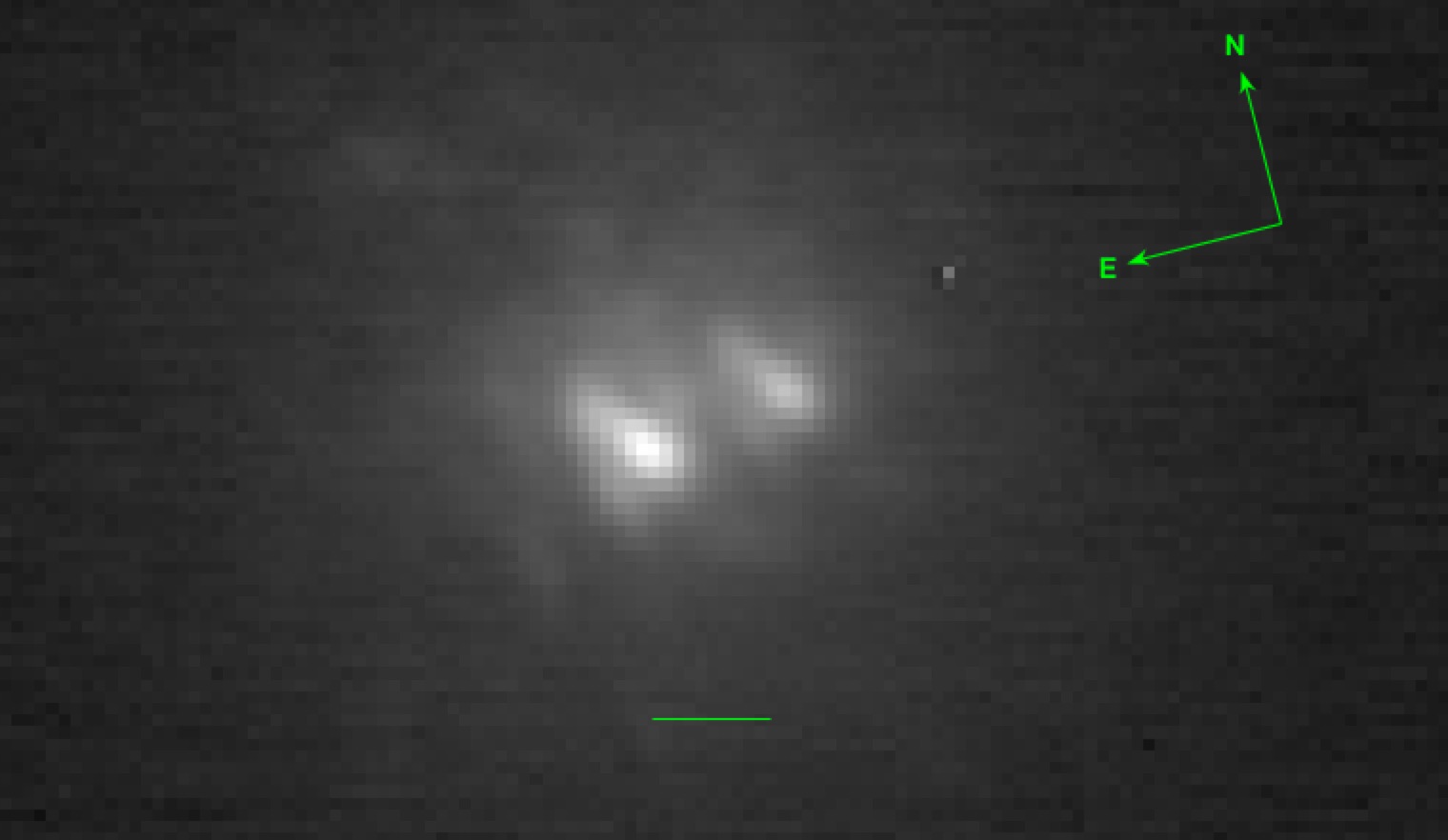} & \includegraphics[width=.48\linewidth]{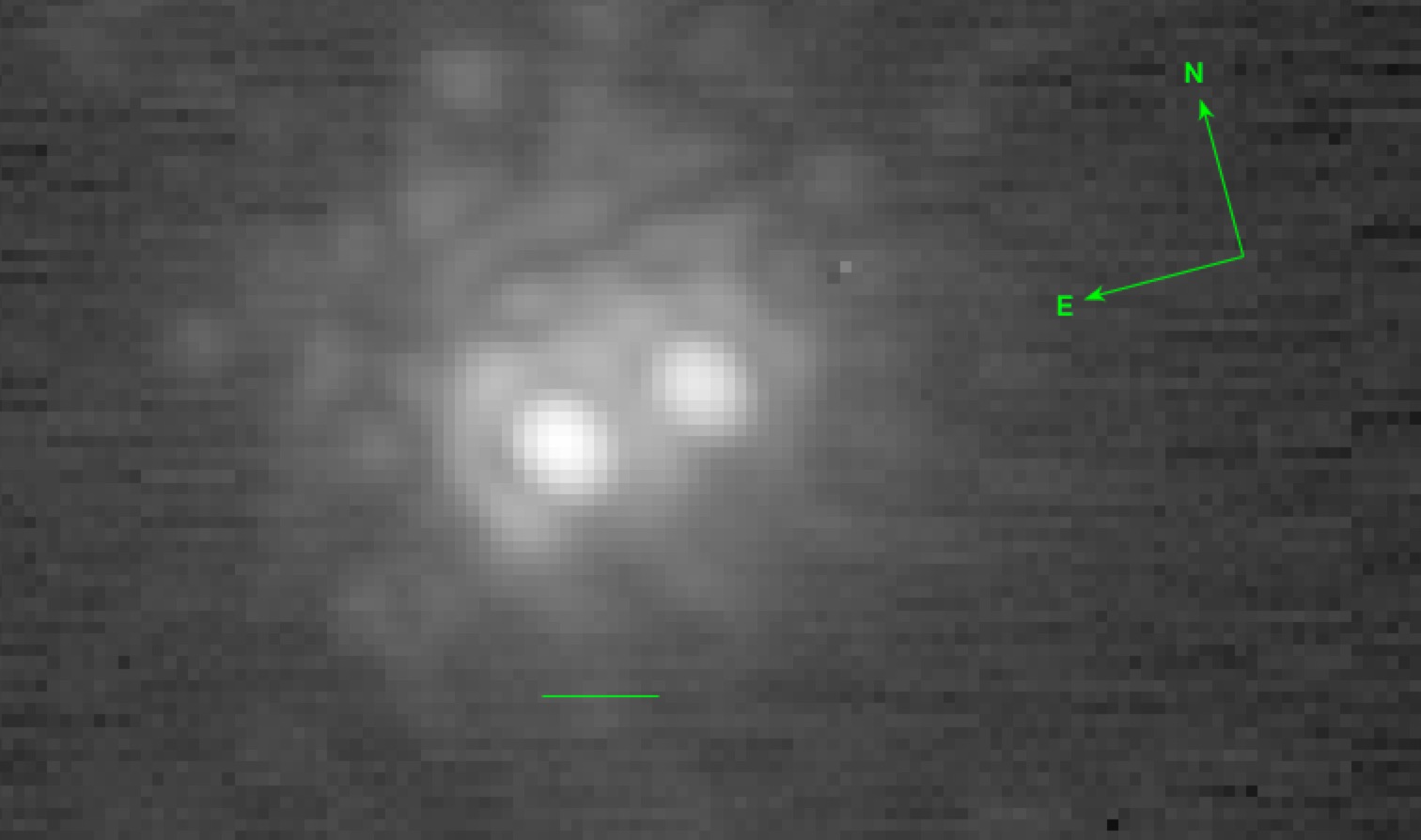} \\

  \end{tabular} \qquad
  \caption{AO images for EPIC 203868608 in $J$ (left column) and $K_P$ (right column) band. Images are shown in logarithmic scale. Compass is shown to indicate north and east. The horizontal green bar in each plot shows 0.1$^{\prime\prime}$ scale. Top row show images when B (the west component) is not eclipsed and bottom row shows images when B is during the primary eclipse. \label{fig:ao}}
\end{figure*}

The raw data were processed using a standard procedure including replacing bad pixels, subtracting dark frames, flat-fielding, and subtracting sky background. We constructed a bad pixel map using dark frames. Pixels with dark currents that deviated more than 5$\sigma$ from their surrounding pixels were recorded as bad pixels. Their values were replaced with the median flux of the surrounding pixels. Dark frames were obtained with the exact same setting as the science frames, e.g., exposure time, co-adds, and readout mode. After dark subtraction, each science frame was corrected for flat fielding. The reduced AO images (shown in Fig. \ref{fig:ao}) were later used for photometric and astrometric measurements. 

\subsection{Keck/HIRES}
\begin{figure*}
\epsscale{1.0}
\plotone{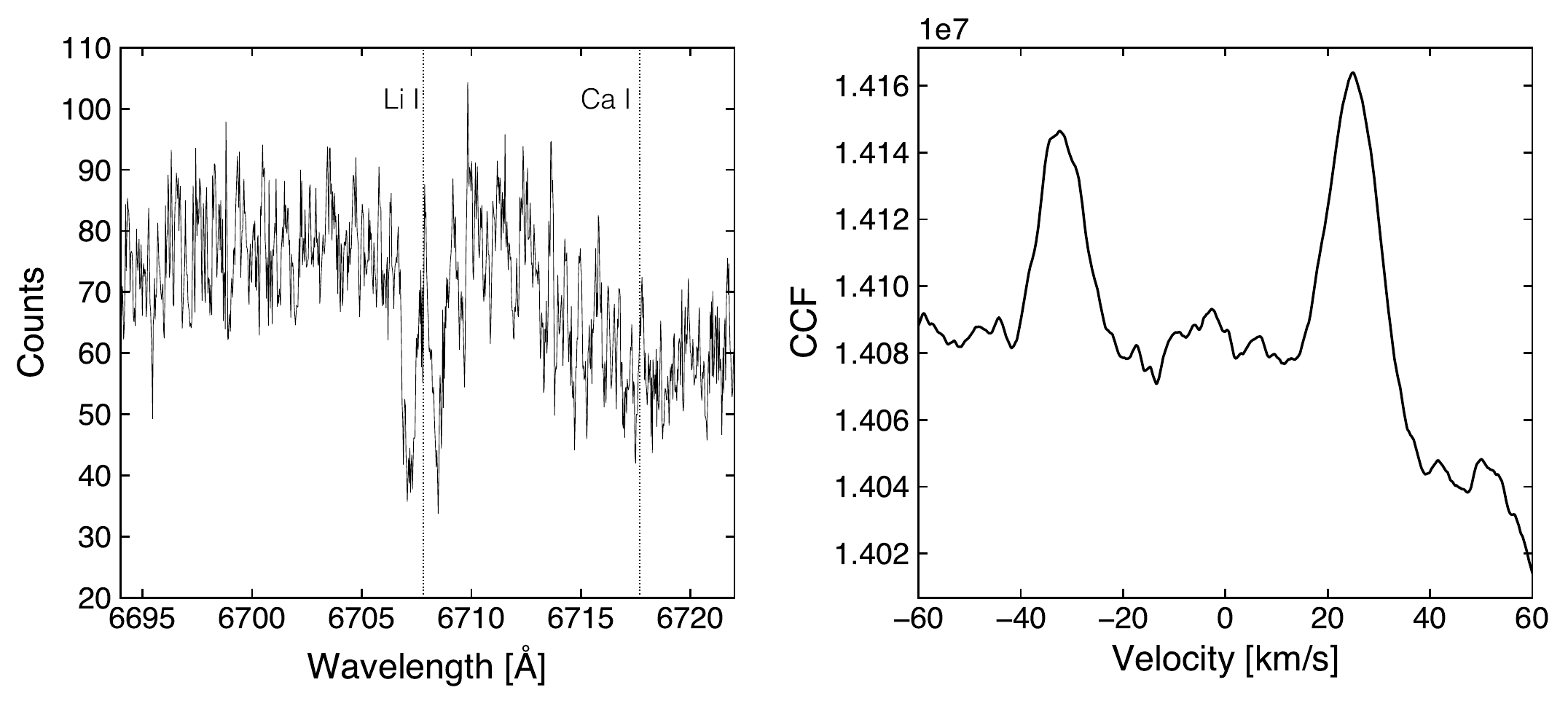}
\caption{Left: a section of the HIRES spectrum for EPIC 203868608 at an epoch when the SB2 is near quadrature (2016 May 20). The Li I 6707.8 $\AA$ line is resolved for each component. The HIRES spectrum contains light from both the SB2 and EB, which are spatially unresolved. The EB component is too faint to contribute measurable features in HIRES spectra. Right: results of a cross-correlation of the spectrum at left with another HIRES spectrum of EPIC 203868608 taken near conjunction of the SB2 orbit. Two peaks of the cross correlation function are clearly visible, indicating two SB components. 
\label{fig:HIRES_SB}}
\end{figure*}

We obtained high dispersion spectra for EPIC 203868608 using Keck/HIRES~\citep{Vogt1994} at 13 epochs between June 2015 and July 2017. From the Keck/HIRES spectra we determined radial velocities (RVs) for the brighter components in the multiple stellar system. The majority of our data were acquired using the B2 or C5 deckers that provide spectral resolution of 70,000 or 36,000, respectively, in the wavelength range from 4800 \AA~to 9200 \AA. In this work, we also include previously published radial velocities derived from HIRES spectra acquired using the setup of the California Planet Search, covering $\sim$3600-8000 \AA~at R$\sim$48,000 with the C2 decker. An example of HIRES spectra centering on Li I 6707.8 $\AA$ line is given in Fig. \ref{fig:HIRES_SB}, showing the SB2 nature of  EPIC 203868608 A. While some of the HIRES RVs are published in D16, the remainder will be presented in David et al. 2018 (\textit{in prep.}). 

\subsection{Keck/NIRSPAO}
\subsubsection{Instrument Setup}
We observed EPIC 203868608 using Keck/NIRSPAO mode in $K$ band, for which the filter was ``NIRSPEC-7-AO". We selected a slit with width of 0.041$^{\prime\prime}$ and length of 2.26$^{\prime\prime}$. The slit width corresponds to a 3 pixel sampling on detector. The spectral resolution is $\sim$25,000 for the slit width.

\subsubsection{Observation}
We observed EPIC 203868608 on UT 2017 Jul 06, coinciding with the primary eclipse of the EB. We started to take data at UT 06:30 and finished taking data at UT 09:22. We used the ``ABBA" dither pattern. Seeing was between 0.8$^{\prime\prime}$ and 1.0$^{\prime\prime}$. Wind speed was low between 0 and 5 mph. 

Exposure time was set to be 600 seconds and 1 coadd per frame. The exposure was chosen to be short enough to resolve the eclipse duration (i.e., a few hours), and long enough for a decent signal to noise ratio (SNR). The peak flux recorded on the detector was $\sim$80-120 ADU (gain = 5.8 $e^{-1}$ per ADU) depending on target airmass and seeing conditions. 

We obtained 4 ABBA patterns corresponding to a total on-target time of 2.67 hours. Compared to the wall time duration of 2.87 hours, the observing duty cycle was 93\%.

\subsubsection{Reducing NIRSPAO Data}
We reduced the NIRSPAO data with a python package \textsc{PyNIRSPEC}~\citep{Boogert2002, Piskorz2016}. The procedures were as follows. Dark frames were subtracted from the raw images which were then flat fielded. Bad pixels were identified in dark frames and their values were replaced by interpolating values of the surrounding pixels.  

The raw images were then divided into different orders. Each order was processed independently including the following procedures: rectification and wavelength calibration. The details of data reduction can be found in ~\citet{Wang2017b}. The final data products of \textsc{PyNIRSPEC} are wavelength-calibrated and rectified 2-d spectra.

We then extracted 1-d spectra from the 2-d spectra. The procedure was complicated for the EPIC 203868608 case: the two visual components were separated by 0.126$^{\prime\prime}$, which is $\sim3\lambda$/D in $K$ band. It is therefore expected that the extracted 1-d spectrum of each visual component was contaminated by the other component. We describe our approach to minimize and remove the flux contamination as follows. 

We created master spectra for A and B by stacking individual spectra over the course of observation. To minimize flux contamination, we only used the half of the point spread function (PSF) that was away from the other component to extract the 1-d spectrum. We used a 2-component Moffat function to model the PSF. The master spectra for A and B were later used to remove flux contamination. 

We then extracted 1-d spectra at different epochs. The more points along the PSF were used, the better the SNR. However, a larger contamination was incurred when more points along the PSF were used. We set quantitative criteria to decide which points are used for 1-d spectral extraction. First, the signal needed to be at least two times higher than the contamination. Second, the signal needed to be higher than 1/20 of the peak signal. The criteria ensured that the flux contamination was always smaller than 15\% without significantly sacrificing incoming signal. 

\begin{figure}
\epsscale{1.1}
\plotone{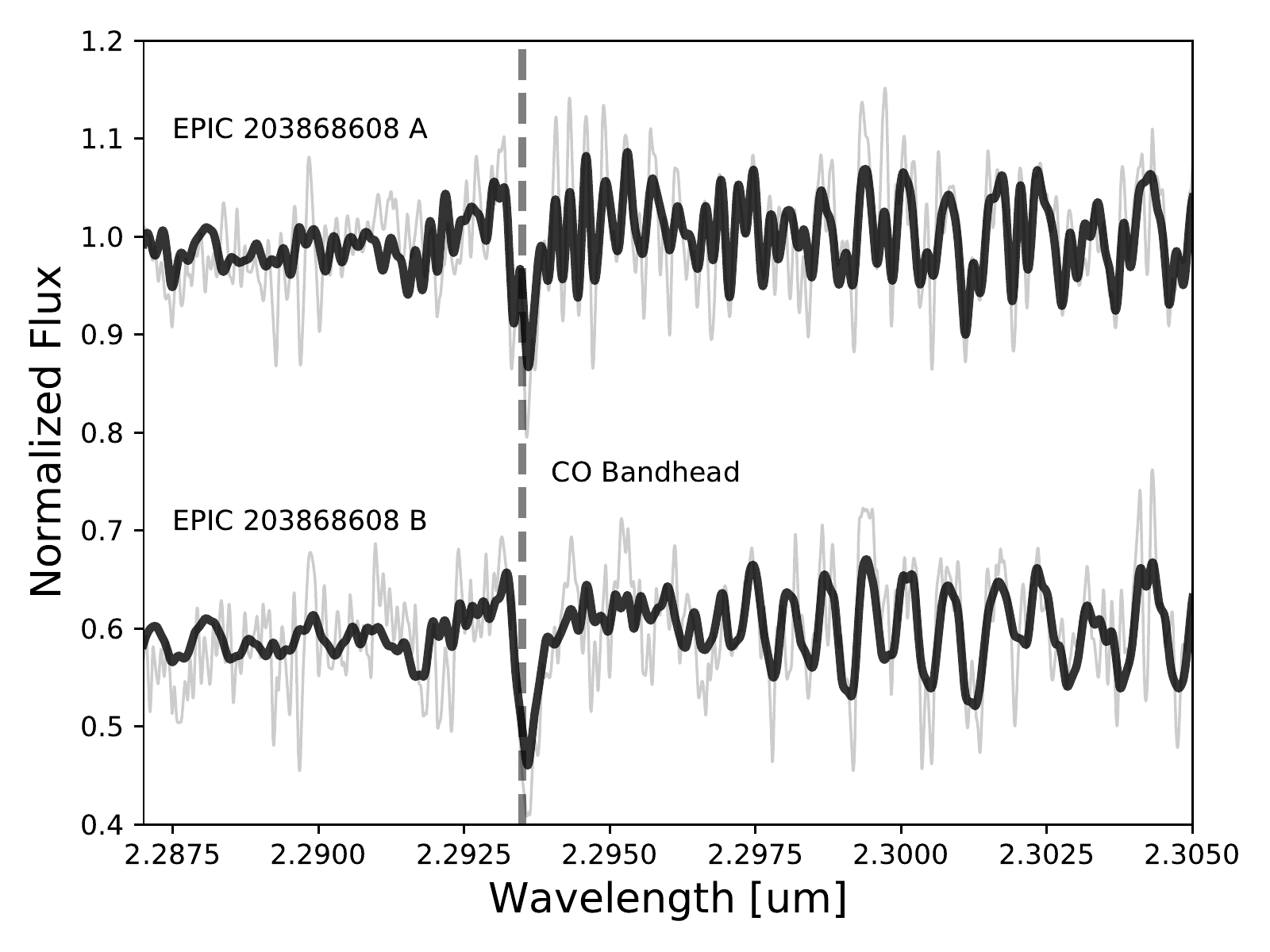}
\caption{A section of NIRSPAO spectra for spatially-resolved EPIC 203868608 A (top) and B (bottom) around a CO bandhead at 2.2935 $\mu$m. The two normalized spectra are offset for visual clarity. CO bandhead is marked as a vertical dashed line. Grey lines are measured spectra, and black lines are a synthetic spectrum convolving with kernels determined by least square deconvolution (\S \ref{sec:lps}). Two sets of spectra can be seen in A spectrum, indicating the SB2 nature of EPIC 203868608 A.
\label{fig:NIRSPAO_EB}}
\end{figure} 

To further decrease the contamination level, we modeled the contamination and removed it from the extracted 1-d spectrum. The contaminating spectrum was obtained as the master spectrum for each component. The contamination level was calculated based on the 2-component Moffat model. We integrated over the pixels that were used in the spectral extraction for the signal PSF and the contamination PSF. The ratio between the two integrals was the contamination level. We then removed the contamination from the single-epoch 1-d spectrum.  

For each visual component, we combined decontaminated spectra for each ABBA pattern, which resulted in spectra at 4 epochs. For the A and B dither position, spectra were shifted along both the slit direction and the dispersion direction. Therefore, we shifted spectra so that they aligned in wavelength space. To do so, we cross-correlated spectra from the A and B detector positions, found the wavelength offset, and then aligned the spectra. 

\section{Results}
\label{sec:results}

\subsection{Orbital Architecture}
\begin{figure}
\epsscale{0.9}
\plotone{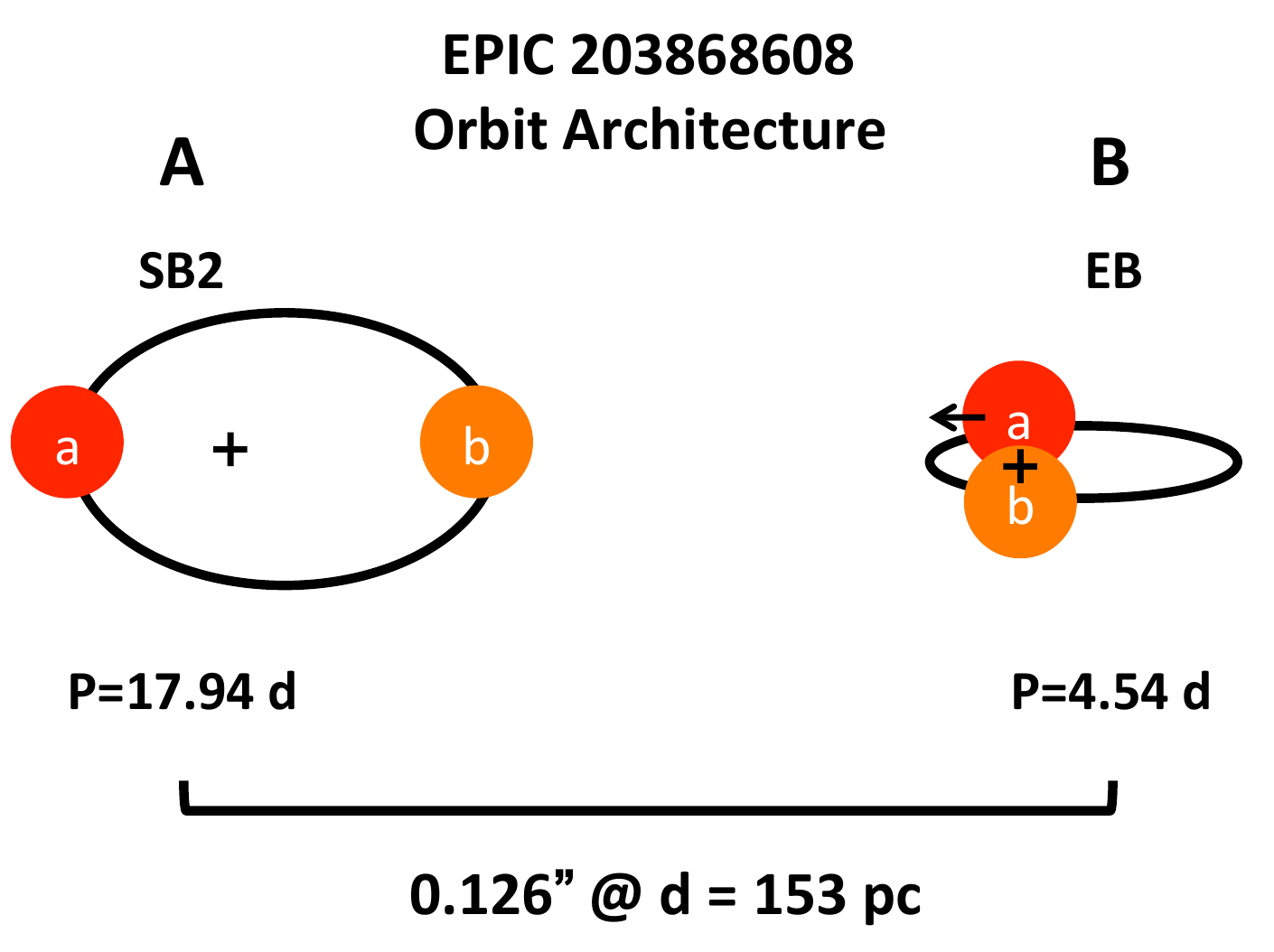}
\caption{Illustration of orbit architecture for EPIC 203868608. The system consists of two visual components (A and B) that are separated by 0.126$^{\prime\prime}$. A is a spectroscopic binary on an eccentric orbit with a period of 17.94 days. The plus sign marks the focus of the ecliptic orbit. B is an eclipsing binary with a period of 4.54 days. The arrow on star a marks the rotational axis. See Table \ref{table:epic608} for more information about the system.
\label{fig:orbit}}
\end{figure}

EPIC 203868608 is a quadruple system that consists of two binary systems separated by 0.126$^{\prime\prime}$ (A and B, see Fig. \ref{fig:orbit}). Aa and Ab compose a SB2. The orbit of component A is mapped out using RVs from HIRES and NIRSPAO (\S \ref{sec:rv}). Ba and Bb compose an EB, whose orbital period is 4.54 days. The diluted light curve is measured by \textit{K2} photometry and a solution for the orbital elements was presented in D16. The EB nature is also confirmed by Keck NIRC2 photometry measurements for in and out of primary eclipse (\S \ref{sec:ao}). At a distance of 153 $\pm$ 7 pc \citep{GaiaDR2}, A and B have a projected separation of 19.3 AU. This separation corresponds to an orbital period of more than 80 years assuming a total mass of 1 solar mass for the quadruple system. Current astrometric data are not adequate to constrain the orbit for A and B. The fact that component A is not eclipsing in \textit{K2} photometry indicates that the orbital planes of A and B are not strictly co-planar at the present time.

\subsection{Radial Velocities}
\label{sec:rv}

Radial velocities for the non-eclipsing SB2 were determined from the Keck/HIRES spectra via cross-correlation with RV standards using the FXCOR task in IRAF\footnote{IRAF is distributed by the National Optical Astronomy Observatory, which is operated by the Association of Universities for Research in Astronomy (AURA) under a cooperative agreement with the National Science Foundation}. FXCOR uses the \cite{Tonry:Davis:1979} cross-correlation method and Gaussian or parabolic profiles to interactively fit for velocity shifts between the two components. We chose orders with high spectroscopic information content that exhibited the highest S/N and were relatively free of significant telluric contamination to determine RVs. For each component, we used the error-weighted means of RV measurements from many individual orders as the final RV. Although the EB and the SB2 are near-equal brightness at NIR wavelengths and both are within the HIRES slit at each epoch, we detected only two clear peaks in the CCF. This is likely due to large $v\sin i$ and fainter magnitude in the optical wavelengths for the EB. The values of the HIRES RVs are presented in (David et al. 2018, \textit{in prep.}).

\begin{figure}
    \centering
    \includegraphics[width=0.9\linewidth]{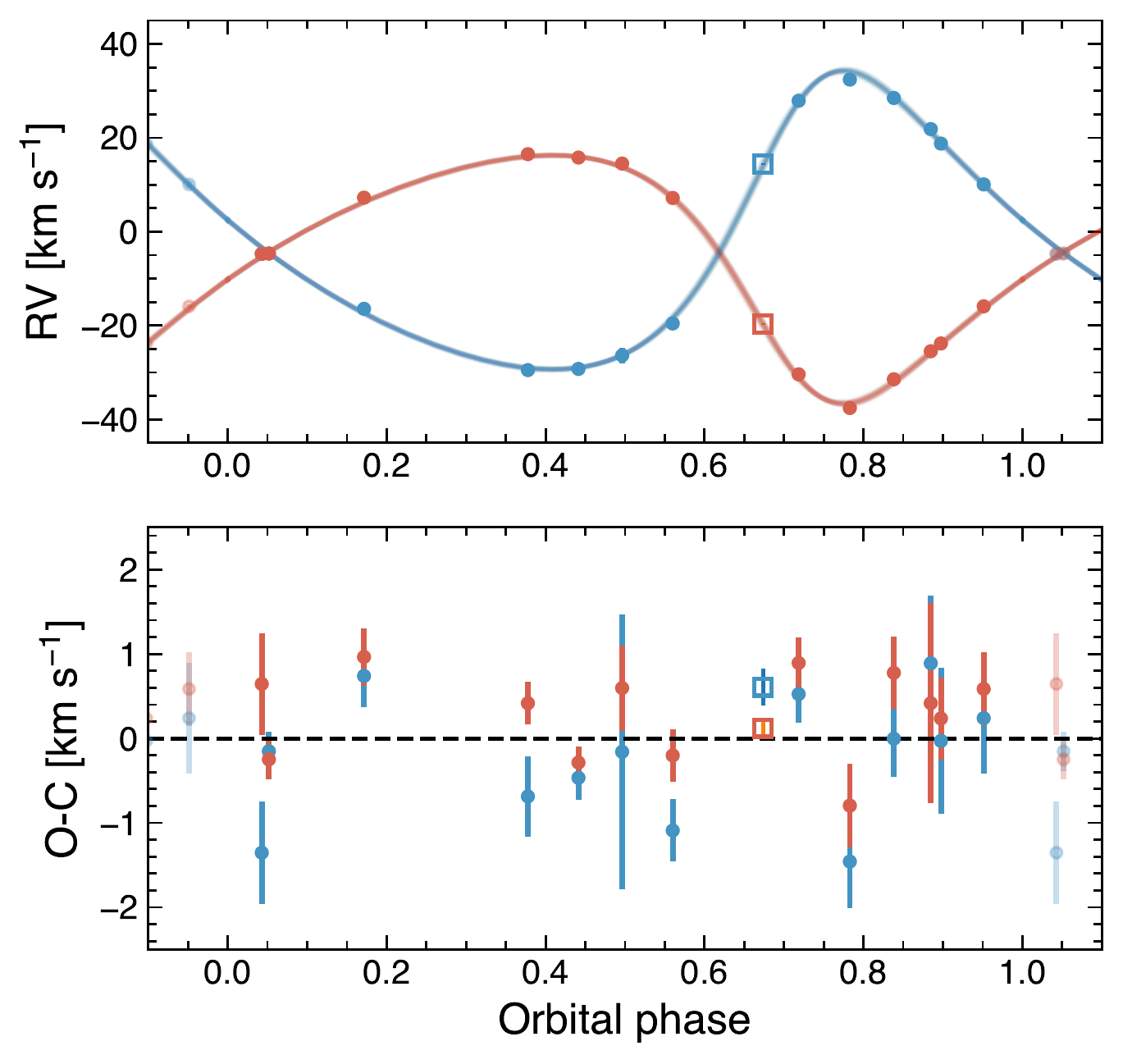}
    \caption{Joint fits to the radial velocity time series of the spectroscopic binary component of EPIC 203868608. Filled circles represent HIRES measurements while the open squares indicate the NIRSPAO measurements. 
    }
    \label{fig:epic608_sb2_fit}
\end{figure}

With the \textsc{jktebop} software we performed joint fits to the RV time series in order to determine  orbital and physical parameters of the SB2. We present these parameters in Table~\ref{table:epic608sb2}, where the uncertainties were determined from 10,000 Monte Carlo simulations. We show fits to the RV time series in Figure~\ref{fig:epic608_sb2_fit}. We find a minimum system mass of $(M_{Aa}+M_{Ab})\sin^3i=0.3685\pm0.0050$~\msun. If one assumes the expected value of $\left \langle \sin^3i \right \rangle = 3\pi/16$, this translates to a system mass of $(M_{Aa}+M_{Ab})\sim0.63$~\msun. Further details on the modeling of the RVs can be found in David et al. (2018, \textit{in prep.}), which supersedes the D16 work that was based on the assumption (now realized as erroneous) that the EB and the SB2 period were the same, rather than arising from two different orbit signals.

\subsection{Photometry and Astrometry}
\label{sec:ao}
We measured differential photometry for the two visual components in EPIC 203868608 {{using a customized code}}. During the primary eclipse, $\Delta J$ and $\Delta K_P$ were $0.44\pm0.04$ and $0.58\pm0.01$. In another epoch, out of an eclipse,  $\Delta J$ and $\Delta K_P$ were $0.23\pm0.01$ and $0.28\pm0.01$. We used a photometric aperture of 6 pixels to measure the flux for A and B. We used an annulus with a radius of 60 pixels and width of 20 pixels to estimate the background. The measurement uncertainty was calculated using the standard deviation of measurements for different individual frames. The differential photometry measurement indicates that the fainter visual component (component B) is responsible for the eclipses observed in K2 photometry and presented in D16. 

{{Astrometric measurements were also conducted with a customized code, in which stellar PSFs were fitted by a 2-d Gaussian function. The centroids of the fits were used to calculate angular separation and position angle between A and B. We used a pixel plate scale of 9.952 mas~\citep{Yelda2010}.}} The measurements were consistent between the two epochs and between the two filters. The angular separation between A and B is $0.126^{\prime\prime}\pm0.004^{\prime\prime}$. The position angle of B with respect to A is $-80.99^\circ\pm0.10^\circ$ (see also Table \ref{table:epic608}).

\subsection{System Masses}
\label{sec:masses}

\begin{figure}
    \centering
    \includegraphics[width=0.95\linewidth]{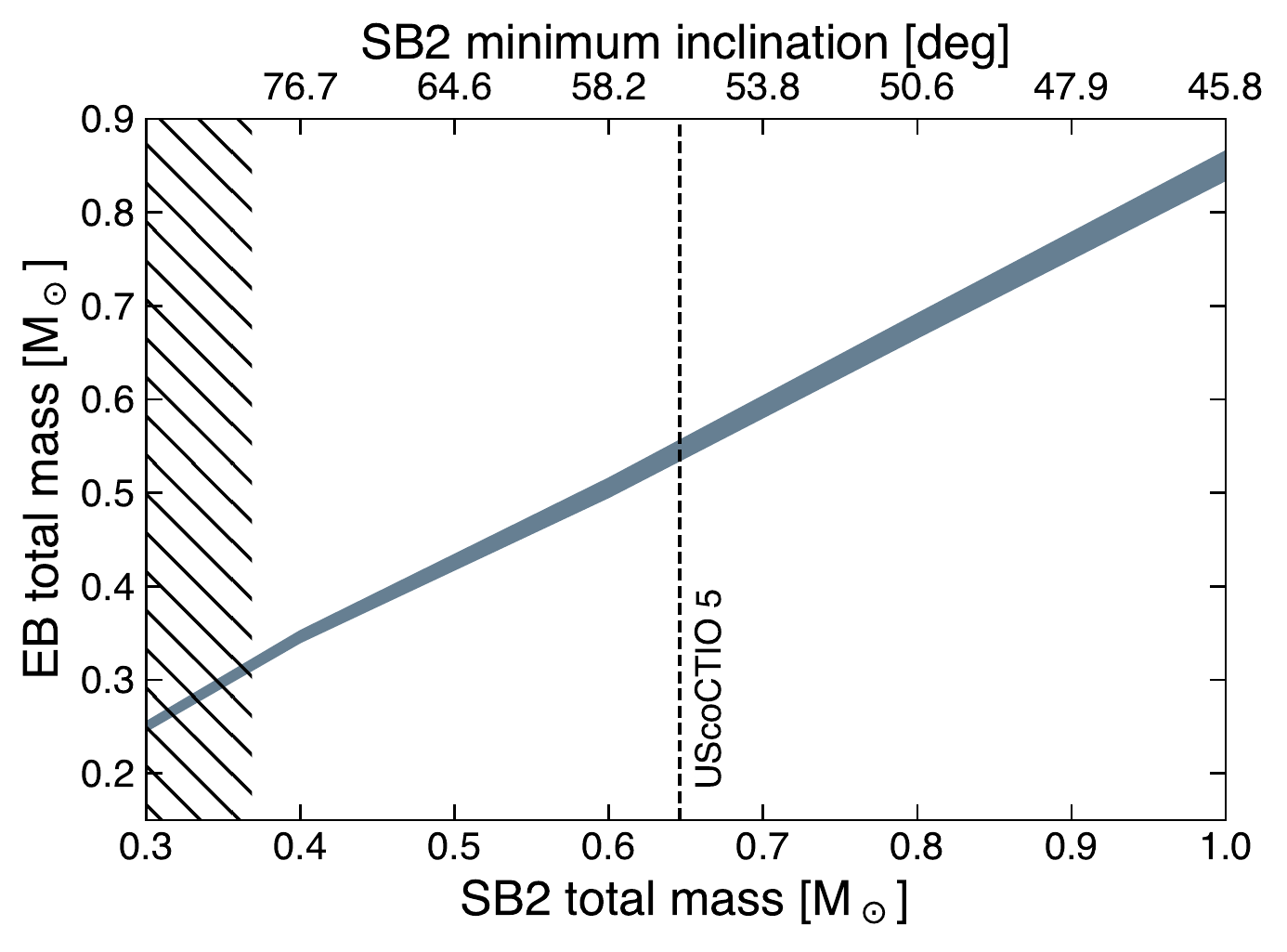}
    \caption{Plausible masses for the spectroscopic binary EPIC~203868608~A and the eclipsing binary EPIC~203868608~B based on BHAC15 models and the $J$ and $K_P$ band contrasts, indicated by the narrow diagonal shaded band. The left-most region (hatched region) of parameter space is excluded by the minimum mass of the SB2. The vertical dashed line indicates the total mass of the young binary UScoCTIO 5, which has a similar primary spectral type and is a binary with well-determined parameters. We consider solutions to the right of this line to be highly unlikely for the now-appreciated quadruple star system, but can not rule them out.}
    \label{fig:epic608_masses}
\end{figure}

From the out-of-eclipse contrasts in the $J$ and $K_P$ bands and the \citet{Baraffe2015} models, we calculated plausible system masses for the EB given a range of assumed system masses for the SB2 (Fig.~\ref{fig:epic608_masses}). In this analysis we assumed for simplicity that both components of the SB2 are equal in mass, and both components of the EB are equal in mass. We assumed an age of 8 Myr, though this analysis is fairly insensitive to the choice of age between 5 and 10 Myr. Solutions in which the SB2 are less than the minimum mass measured from orbit-fitting could be excluded. We also considered solutions in which the total mass of the SB2 is greater than the total mass of the binary UScoCTIO~5 \citep[total mass of 0.65~\msun, ][]{Kraus2015,David2016} to be highly unlikely. This is because EPIC~203868608 clearly resides at a fainter and redder position in the color--absolute magnitude diagram with respect to UScoCTIO~5, which has only two components despite sharing a similar spectral type. We thus considered the most plausible mass range for the SB2 to be $\sim$0.4--0.6~\msun, corresponding to a range of $\sim$0.3--0.5~\msun for the EB.


Due to the fact that only the SB2 is detected in our Keck/HIRES spectra, we can not presently measure fundamental masses and radii for the EB. Future efforts using either high-resolution IR spectroscopy or spatially-resolved spectroscopy at a range of phases should allow for the determination of masses and radii (A. Kraus, private communication). Moreover, long-term astrometric monitoring via high-resolution imaging should enable a determination of the wide EB+SB2 orbit. At that point, it should be possible to determine dynamical masses for each of the four components in the system and begin to place stringent constraints on evolutionary models for low-mass stars \citep[e.g.][]{Mathieu1994, Mathieu2007, Hillenbrand2004, Stassun2014}.

\subsection{Comparing EPIC 203868608 to Other Quadruple Systems}
The orbital architecture of EPIC 203868608 is typical of other known quadruple systems, which regularly show four stars with similar masses and similar periods of the inner subsystems \citep{Tokovinin2008}. The $\epsilon$ Lyr system (composed of 4 A-type stars) is considered a prototype of this type of quadruple architecture, and BD -22 5866 \citep{Shkolnik2008} is an example at masses similar to those in the EPIC 203868608 system. One proposed formation channel for such hierarchical quadruples is through cascade fragmentation, in which a rotating core collapses into a centrifugally-supported disk which then undergoes further rotational fragmentation if the angular momentum of fragments in the disk is high enough \citep{Bodenheimer1978}. \textit{N}-body dynamics between fragments within a pre-stellar core is another proposed mechanism for the origin of such 2+2 quadruples or ``double twins,'' although such an outcome is rare with this mechanism \citep{Delgado2004}. Assuming an outer period of 80 years, the location of EPIC 203868608 in a $P_\mathrm{in}-P_\mathrm{out}$ diagram, where $P_\mathrm{in}$ is the period of either of the inner subsystems (comparable in this case), is indeed quite close to the region of highest density among known quadruple systems \citep[see Figure 11 of][]{Tokovinin2008}. This observation would seem to suggest that whatever mechanism is responsible for creating such systems would need to act on timescales much shorter than that of the age of EPIC 203868608 ($\lesssim$10~Myr).

EPIC 203868608 joins a relatively small list of pre-main-sequence (PMS) quadruple systems which, if characterized well, can place tight constraints on evolutionary models and formation scenarios. Other notable PMS quadruples include GG Tau \citep{White1999}, V773 Tau \citep{Boden2007}, LkCa 3 \citep{Torres2013}, and 2M0441+2301 \citep{Bowler2015}.

\section{EB Obliquity}
\label{sec:oblqt}
Obliquity is defined as the angle between the rotational axis of the eclipsed star and the orbital angular momentum vector of the eclipsing system. The obliquity for an EB system can be measured via the Rossiter-McLaughlin (RM) effect~\citep{Rossiter1924, McLaughlin1924} or, more generally, through spectral line profile (LP) changes during an eclipse, i.e. Doppler tomography.

For a fast-rotating object, the spectral LP is mainly broadened by the rotation. During an eclipse, the spectral LP would deform because certain velocities are missing due to the occultation. As the eclipse progresses, the spectral LP deformation exhibits certain patterns for a given obliquity. For example, the spectral LP would red-shift and then blue-shift for a pro-grade orbit and vice versa. In subsequent subsections, we will describe the procedure to measure LPs during an eclipse (\S \ref{sec:lps}), model LPs (\S \ref{sec:lpm}), and infer the EB obliquity from the LP measurement and modeling (\S \ref{sec:lpr}). 

\subsection{LP Measurement}
\label{sec:lps}
We derived the LP using the least square deconvolution method. The method is detailed in~\citet{Wang2017b} and can be summarized by the following equation:
\begin{equation}
\label{eq:lsd}
\mathbf{Z=(M^T\cdot S^2\cdot M+ R)^{-1}\cdot M^T\cdot S^2\cdot Y^0},
\end{equation}
where matrix transpose is denoted by $\mathbf{T}$, $\mathbf{Z}$ is the LP, $\mathbf{M}$ is a $m \times n$ Toeplitz matrix, where $m$ is the number of data points in a spectrum and $n$ is the desired number of data points in the LP. $\mathbf{M}$ is generated from a template spectrum $\mathbf{F}$ that has the same wavelength sampling as the observed spectrum $\mathbf{Y^0}$. $\mathbf{S}$ is an $m \times m$ matrix with $\mathbf{S}_{ii} = 1 / \sigma_i$, where $\sigma_i$ is the measurement error for each spectral data point. {{We use PHOENIX BT-Settl spectrum~\citep{Allard2001} with T$_{eff}$=2900 K and log(g)=4.0 as our template spectrum $\mathbf{F}$. The spectrum gives the least residual in the least square deconvolution.}}

$\mathbf{R}$ is a regularization matrix. We used a first-order Tikhonov matrix as the regularization matrix in~\citet{Wang2017b}. Here, we use the inverse of a covariance matrix as the regularization matrix. We denote the covariance matrix as $\mathbf{K}(\alpha_Z)$, where $\alpha_Z$ is a set of parameters. The covariance matrix can have many forms, but we adopt a commonly used form - the squared exponential covariance matrix:
\begin{equation}
K_{ij}(\sigma_Z, \lambda_Z)=\sigma_Z^2\exp\left[-\frac{(v_i-v_j)^2}{2\lambda_Z}\right],
\end{equation}
where $v_i$ is the corresponding velocity for a LP. 

There is an advantage to replacing the first-order Tikhonov matrix with an inverse covariance matrix: $\mathbf{Z}$ is a multivariate Gaussian distribution that can be calculated in the Bayesian framework~\citep{Ramos2015}:
\begin{equation}
\mathbf{Z} = \mathbb{N}(\mathbf{\mu_Z, \Sigma_Z}), 
\end{equation}
where $\mathbb{N}$ denotes a multivariate Gaussian distribution, $\mathbf{\mu_Z}$ and $\mathbf{\Sigma_Z}$ are the mean and covariance matrix for the multivariate Gaussian distribution. $\mathbf{\Sigma_Z}$ and $\mathbf{\mu_Z}$ can be calculated using the following two equations as derived from~\citet{Ramos2015}:
\begin{equation}
\mathbf{\Sigma_Z} = [\mathbf{M^T\cdot S^2\cdot M}+\mathbf{K}(\hat{\alpha}_Z)^{-1}]^{-1}, 
\end{equation}
where $\hat{\alpha}_Z$ is the set of parameter that maximizes the marginal posterior for $\alpha_Z$. Assuming the distribution of $\alpha_Z$ is strongly peaked, we can follow the Type-II maximum likelihood solution~\citep{Bishop1995} for $\mathbf{\Sigma_Z}$ and $\mathbf{\mu_Z}$. With $\mathbf{\Sigma_Z}$ calculated, $\mathbf{\mu_Z}$ can be calculated using the following equation:
\begin{equation}
\mathbf{\mu_Z = \Sigma_Z\cdot M^T\cdot S^2\cdot Y^0}.  
\end{equation}

\begin{figure*}
  \centering
  \begin{tabular}[b]{cc}
    \includegraphics[width=.4\linewidth]{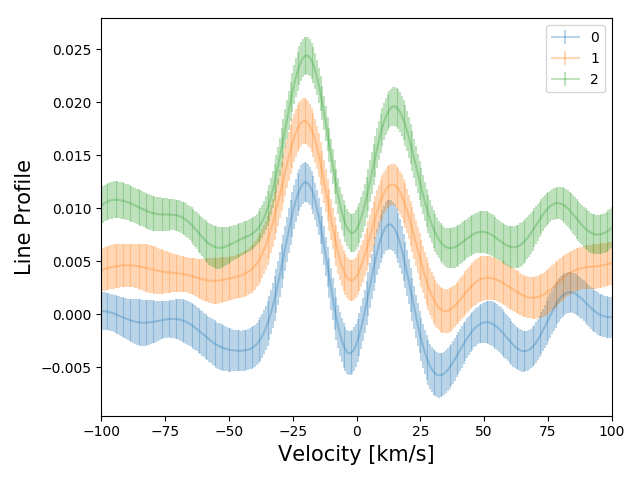} & \includegraphics[width=.4\linewidth]{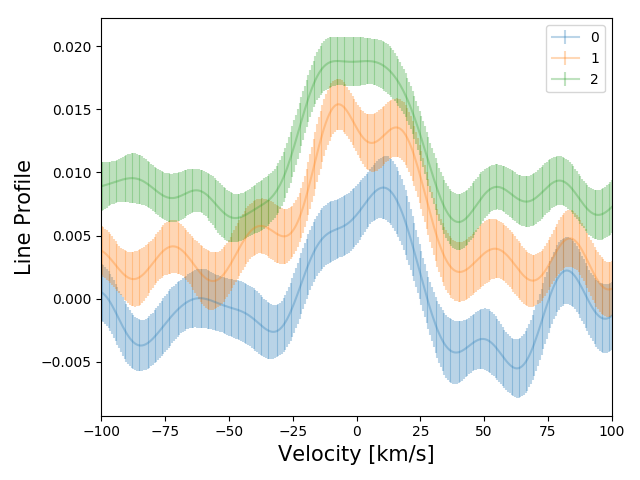} \\
  \end{tabular} \qquad
  \caption{Left: line profiles (LPs) for EPIC 203868608 A at three epochs of NIRSPAO observations. Velocity shift of LPs are corrected for instrument drift (using telluric lines) and barycentric velocity. Right: the same as left but for EPIC 203868608 B. Colors and numbers represent epochs in NIRSPAO observations: 0, mid-eclipse; 1, egress; and 2, out-of-eclipse.  \label{fig:lp_atm_bd}}
\end{figure*}

Fig. \ref{fig:lp_atm_bd} shows LPs for A and B based on NIRSPAO observations. Three epochs are shown for mid-eclipse (-0.08 hour), egress (0.67 hour), and out-of-eclipse (1.32 hour). The times in parentheses indicate elapsing time with respect to the center of the primary eclipse (at 0.0 hour) for an eclipse that lasts for 2.7 hours. Note that we combine the last two epochs of NIRSPAO observations to increase the SNR for the out-of-eclipse LP.

\subsection{RVs for EPIC 203868608 A and B Derived from LP}

Owing to the AO system that spatially separates the two visual components, we can measure RVs for individual components using the following equation: $v=\bar{v}-v_{\rm{atm}}+\rm{bcc}$, where $\bar{v}$ is the velocity center of measured LP, $v_{\rm{atm}}$ is the velocity center of measured LP for telluric lines, bcc is barycentric correction~\citep{Wright2014}. We use two methods to calculate the velocity center of a LP: measuring the flux-weighted centroid and polynomial fitting for the LP peak. Both methods yield consistent results. The RV value and uncertainty are estimated by repeating the RV measurement for 100 iterations. In each iteration, correlated Gaussian noise is added to the LP. We report the average and the standard deviation as the RV value and its uncertainty. 

The RVs for Aa and Ab are $14.39\pm0.22$ km$\cdot$s$^{-1}$ and $-19.71\pm0.12$ km$\cdot$s$^{-1}$, respectively. These are consistent with the orbital solution found from the HIRES RV measurements (see Fig. \ref{fig:epic608_sb2_fit}). We do not resolve the RV difference between Ba and Bb because the observation was taken when Ba and Bb were eclipsing and $\Delta$RV should be around zero. Instead, we measure the systemic RV for B, which is $0.38\pm0.69$ km$\cdot$s$^{-1}$. 

The offset of systemic RV between A ($-4.44\pm0.07$ km$\cdot$s$^{-1}$) and B ($0.38\pm0.69$ km$\cdot$s$^{-1}$) is significant at the 6-$\sigma$ level. This casts doubt on the physical association between A and B, although both velocities are within $\sim$2$\sigma$ of the median radial velocity for the Upper Sco association as a whole \citep{GaiaDR2}. However, the discrepancy may be reconciled by the following counter arguments. First, the orbital motion of B around A could account for up to $\sim$3 km$\cdot$s$^{-1}$ (i.e. a majority of the discrepancy) assuming a total mass of 0.8 \msun~and a separation of 20 AU. Second, the angular separation of 0.126$^{\prime\prime}$ makes it unlikely to have an equal-brightness optical double that is physically unassociated~\citep{Horch2014}. 

\subsection{LP Modeling}
\label{sec:lpm}
We adopt an analytic solution to model LPs for an EB~\citep[][\it{in prep.}]{Pezzato2018}. The analytic model greatly reduces the computational time compared to finite element models that are usually used for EB LP analysis~\citep[e.g., ][]{Albrecht2007, Johnson2014}. Our approach is similar to the analytic method to model the RM effect for transiting planets~\citep{Hirano2011}. The deformed LP during transit is the out-of-transit LP minus a ``Doppler shadow'', which is the line spread function (LSF) of the spectrograph centered at the blocked velocity and scaled by the flux blocked by the planet. The differences are: (1) the occulting object can no longer be treated as a point source and (2) the occulting object is self-luminous in the EB case rather than a dark spot in the planet case. 

To address the point source problem, we calculate the maximum and minimum velocity of the occulted area, and construct a rotationally broadened LP based on the two velocities. The LP is then convolved with the LSF of the spectrograph, which is measured using telluric lines at low airmass. 

The self-luminous problem is addressed by adding another LP that is from the occulting star and that is scaled by the relative flux between the secondary and the primary star. Therefore, the final LP is the unocculted LP of the primary, minus the flux-scaled occulted LP, and plus the flux-scaled LP of the secondary, and then normalized so that the total area under the curve equals to unity. The reader is referred to Appendix \ref{app:comp} for a comparison between the analytic model and a finite-element model~\citep[][and references therein]{Albrecht2014}.

The parameters in our analytic model are: impact parameter, obliquity, eclipse duration (2.7 hours), $v\sin i$ for the primary and secondary stars, radius ratio between the primary and the secondary (1.0), orbital period (4.54 days), total mass (0.3 \msun) and mass ratio (0.8), flux ratio (0.8), quadratic limb darkening parameters (0.8 and 0.1)~\citep{Claret2012}, eccentricity (0.3), argument of periastron (100$^\circ$), time at periastron (2456896.19699 BJD), and a y-axis scaling factor for LPs. The parameters with values indicated in parentheses are fixed in the following inference for obliquity because they are either well-constrained by observations (e.g. D16) or variation of the parameters does not significantly change the obliquity measurement.   
\subsection{Results for Obliquity Measurement}
\label{sec:lpr}
\begin{figure*}
\epsscale{0.9}
\plotone{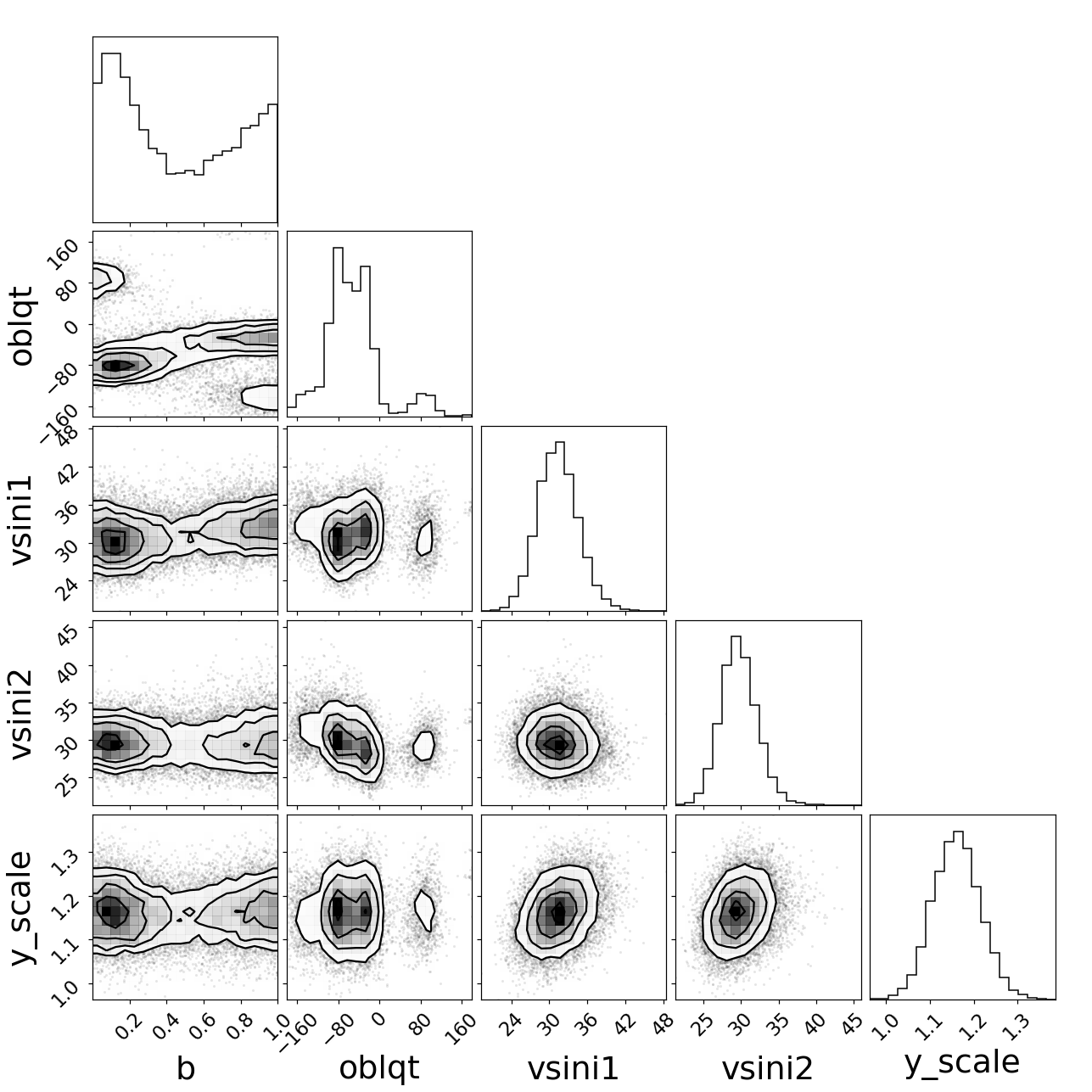}
\caption{Posterior sample distributions for parameters in modeling LP variation during the eclipse of EPIC 203868608 Bab. 
\label{fig:mcmc}}
\end{figure*}

\begin{figure}
\epsscale{0.98}
\plotone{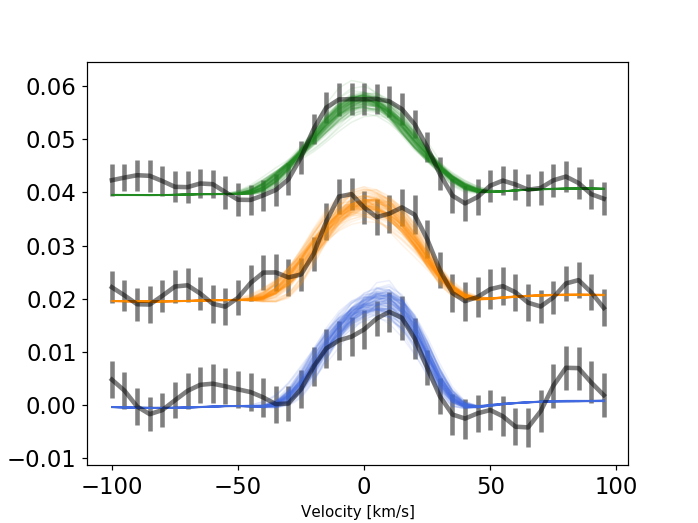}
\caption{Line profiles (LPs) at three epochs: mid-eclipse, egress, and out-of-eclipse. Grey points linked by solid grey lines are LP measurements for EPIC 203868608 Bab (the same as Fig. \ref{fig:lp_atm_bd} Right). Solid colored lines are random draws from MCMC posterior samples.   
\label{fig:LP}}
\end{figure}

We generate posterior samples for the LP parameters by exploring the likelihood space of model fits to the data using an affine-invariant Markov chain Monte Carlo~\citep[MCMC, ][]{Goodman2010} as implemented in the Python package \textsc{emcee}~\citep{ForemanMackey2013}. We include the following parameters in the MCMC: impact parameter (b), obliquity ($\lambda$), and $v\sin i$ for the primary and the secondary star. We apply uniform priors to these parameters within their boundaries, $0<b<1$, $-180^\circ<\lambda<180^\circ$, and $0<v\sin i<50$ km$\cdot$s$^{-1}$. Convergence criteria include: (1) the length of MCMC chains is at least 100 times autocorrelation length and (2) the change of the autocorrelation length estimates between consecutive check are less than 5\%. The chain was considered converged when both of these criteria were met for each freely-fitted parameter. Fig. \ref{fig:mcmc} shows the posterior distribution. {{Median values and uncertainties for different parameters are: $b=0.36^{+0.49}_{-0.26}$, $\lambda=-57^{+40}_{-36}$ degrees, $\rm{V}\sin i_1=31.2^{+3.1}_{-3.0}$ km$\cdot$s$^{-1}$, $\rm{V}\sin i_2=29.6^{+2.4}_{-2.2}$ km$\cdot$s$^{-1}$, and the y-axis scaling factor = $1.16^{+0.05}_{-0.05}$. The uncertainties are calculated by subtracting 68\% interval by the median values. }}Fig. \ref{fig:LP} shows LPs that are drawn from the posterior samples in comparison with LP measurements at different epochs. 

{{Posterior distribution of obliquity $\lambda$ shows clearly multi-modal distribution. While the posterior number density is low, there is an island of positive obliquity values. These represent cases in which the secondary transverses the opposite side of the primary equator with the same impact parameter $b$. On the island of negative obliquity values where number density is the highest, there appears to be a bimodal distribution. However, the lower-obliquity peak around 20$^\circ$ is correlated with higher value of $b$, which are cases for grazing eclipse. This situation is unlikely given the low value of $b$ as inferred from light curve fitting~\citep{David2016}. Therefore, a careful examination of the multi-modal distribution of posterior distribution of obliquity further favors an orbit with high obliquity. }}

The large uncertainty in obliquity is due to the following reasons: (1), data quality is not high because of challenging AO-aided high resolution spectroscopy; (2), we do not have a complete coverage of the eclipse, only data points during the eclipse are taken; (3), we have incomplete knowledge on the orbital parameters; and (4), correlated noise in the
observations is not well understood. The reader is referred to \S \ref{app:sensitivity} for discussions on the sensitivity of our NIRSPAO observations to the RM effect from EPIC 203868608.   

\subsection{Implications of an Inclined Orbit}
\label{sec:imp}
Doppler tomography data suggest that EPIC 203868608 Bb is on an inclined orbit around Ba with an obliquity of $-57^{+40}_{-36}$ degree with a 2-$\sigma$ upper limit at -2.9$^\circ$. 

The inclined orbit has significant implications for the stellar properties of low-mass stars and the formation history of the quadruple system. Specifically, the high obliquity puts constraints on the tidal quality factor Q and the mechanism through which EPIC 203868608 forms. 
\subsubsection{Tidal Quality Factor}

EPIC 203868608 is young and both components have eccentric orbits. This indicates that the tidal circularization time scale should be longer than the system age ($\sim10$ Myrs). Here we use that fact to constrain the tidal quality factor for the low-mass stars in EPIC 203868608. Since constraints are stronger for systems with shorter periods, we will discuss the EB component, the period of which (P=4.54 days) is shorter than that for the SB2 (P=17.94 days). The rate at which eccentricity changes due to tidal dissipation is given as follows~\citep{Hut1981}:
\begin{equation}
\begin{split}
\label{eq:edot}
\dot{e} = & 27\frac{k}{T}q(1+q)\left(\frac{R}{a}\right)^3\frac{e}{(1-e^2)^{13/2}}\times\\
& \left(f_3(e^2)-\frac{11}{18}(1-e^2)^{3/2}f_4(e^2)\frac{\Omega}{n}\right),
\end{split}
\end{equation}
where $k=0.28$, which is twice the Love number~\citep{Batygin2013}; $q$ is mass ratio, which is assumed to unity here; $R/a$ is the ratio of radius of the primary star to semi-major axis, which is constrained by light curve fitting in D16: $R/a=0.0679$; $e=0.3227$ is eccentricity (D16). The expression of $f_3$ and $f_e$ can be found in~\citet{Hut1981}. $\Omega$ and n are the angular velocities of the rotation and orbit: $\Omega$ is inferred from $v\sin i$ measurement and n is inferred from orbital parameters from light curve fitting in D16. $T$ in Equation \ref{eq:edot} is:
\begin{equation}
\label{eq:T}
T=\frac{R^3}{GM\tau},
\end{equation}
where $\tau$ is the inverse of the product of the tidal quality factor Q and n~\citep{Peale1999}. 

Next, the tidal circularization time scale $\tau_{\rm{circ}}$ can be calculated as $e/\dot{e}$. In order for $\tau_{\rm{circ}}$ to be longer than 10 Myrs, Q needs to be higher than $5\times10^4$. This finding is consistent with previous works on tidal Q~\citep[e.g., ][]{Matsumura2008}. 

Additional constraints on tidal Q can be obtained from the inclined orbit, which suggests that the synchronization process is not finished. To synchronize the rotational and orbital period, the system needs to be aligned in the first place. Therefore, the age of the system should be smaller than the tidal synchronization time scale, which can be estimated using the following equation~\citep{Rasio1996b}:
\begin{equation}
\label{eq:sync}
\tau_{\rm{sync}}\approx Q\cdot\omega\cdot q \left(\frac{R_2^3}{Gm_2}\right)\left(\frac{a}{R_2}\right)^6,
\end{equation}
where $\omega$ is the difference of angular frequency between the primary rotation and orbit, and $R_2$ is the radius of the secondary star. Given that $\tau_{\rm{sync}}>10$ Myrs, Q needs to be higher than $5\times10^5$. Note that the constraint from tidal synchronization is 10 times stronger than that from tidal circularization. Nonetheless, the allowed range for the tidal quality factor (Q $>5\times10^5$) is still consistent with previous studies. 

\subsubsection{Kozai-Lidov Perturbation vs. Stochastic Processes}
We investigate here if the EB was formed through Kozai-Lidov (KL) perturbations~\citep{Kozai1962, Lidov1962}. KL perturbations are known to result in highly-inclined orbit for exoplanets~\citep[e.g., HD 80606,][]{Pont2009}. In order for the KL mechanism to operate, the KL timescale needs to be shorter than the general relativity (GR) precession time scale. The KL time scale can be calculated using the following equation~\citep{Fabrycky2007}:
\begin{equation}
\label{eq:kl}
\tau_{\rm{KL}}=\frac{2\rm{P}_{\rm{out}}^2}{3\pi\rm{P}_{\rm{in}}}\frac{m_1+m_2+m_3}{m_3}(1-e_{\rm{out}}^2)^{3/2},
\end{equation}
where $P$ is orbital period, subscripts in and out denotes the EB and the perturber, $m$ is mass, subscripts 1, 2, and 3 denote primary and secondary in the EB, and the perturber, and $e$ is eccentricity.  

The GR precession time scale can be calculated using the following equation~\citep{Fabrycky2007}:
\begin{equation}
\label{eq:gr}
\tau_{\rm{GR}}=\frac{2\pi c^2}{3\rm{G}^{3/2}}\frac{a_{\rm{in}}^{5/2}}{(m_1+m_2)^{3/2}}(1-e_{\rm{in}}^2),
\end{equation}
where $c$ is the speed of light, G is the gravitational constant, a$_{\rm{in}}$ is the semimajor axis for the EB, and $e_{\rm{in}}$ is the eccentricity for the EB. 

Assuming $\rm{P}_{\rm{in}}=4.54$ days, $\rm{P}_{\rm{in}}=88.6$ years, m$_1=0.15$ \msun, m$_2=0.15$ \msun, m$_3=0.4$ \msun, a$_{\rm{in}}=0.0359$ AU, and $e_{\rm{in}}=0.32$, the condition that $\tau_{\rm{KL}}<\tau_{\rm{GR}}$ is $e_{\rm{out}}>0.82$. However, one needs to note that $\tau_{\rm{GR}}$ is a strong function of a$_{\rm{in}}$. Tracing back in time when a$_{\rm{in}}$ was larger, $\tau_{\rm{GR}}$ can be longer than $\tau_{\rm{KL}}$, and the requirement for $e_{\rm{out}}$ is relaxed to lower values. 

Alternatively, the orbital architecture of EPIC 203868608 may be an evolutionary consequence of an even higher-order multiple star system~\citep{Ghez1993}. It is shown in numerical simulations that hierarchical multiples can result from dynamical interactions of young stellar multiples formed by fragmenting cloud~\citep{Sterzik1998}. Spin-orbit misalignment may take place during this formation stage. In addition, star-disk interaction in a multiple star system can also lead to spin-orbit misalignment~\citep{Spalding2014}.





\begin{deluxetable*}{llcccc}
\tablecaption{Summary of observations \label{table:obs_summary}}
\tablecolumns{5}
\tablewidth{-0pt}
\tabletypesize{\scriptsize}
\tablehead{
        \colhead{Instrument} &
        \colhead{UT Date} &
        \colhead{Wavelength} &
        \colhead{Exposure} &
        \colhead{Configuration} &
        \colhead{Seeing} 
        }
\startdata
\multirow{6}{*}{NIRC2} & \multirow{2}{*}{2015 Jun 22$^{1}$} & $J$ & 40 sec & narrow & \multirow{2}{*}{0.6$^{\prime\prime}$} \\
 & & $K_{P}$ & 40 sec & narrow &  \\
 & \multirow{2}{*}{2015 Jul 25$^{2}$} & $J$ & 90 sec & narrow & \multirow{2}{*}{0.3$^{\prime\prime}$} \\
 & & $K_{P}$ & 300 sec & narrow &  \\
 & 2016 Jul 17 & $K_{P}$ & 60 sec & narrow & 0.4$^{\prime\prime}$ \\
\multirow{2}{*}{HIRES$^{3}$} & 2015 Jun & 3600 & \multirow{2}{*}{varied} & B2, C2 and & \multirow{2}{*}{varied} \\
 & 2017 Jul & -9200 \AA & & C5 decker & \\
NIRSPAO & 2017 Jul 06 & $K$ & 2.67 hours & AO & 0.9$^{\prime\prime}$ 
\enddata
\tablecomments{1, Data from Keck Observatory Archive (KOA), PI: Mann. 2, Data from KOA, PI: Baranec. 3, See~\citet{David2016} for more details.  
}
\end{deluxetable*}

\begin{deluxetable*}{lccc}
\tablecaption{Photometry, astrometry, and orbital and physical parameters of EPIC 203868608 \label{table:epic608}}
\tablecolumns{4}
\tablewidth{-0pt}
\tabletypesize{\scriptsize}
\tablehead{
        \colhead{Parameter} &
        \colhead{Units} &
        \multicolumn{2}{c}{{Value}} 
        }
        
\startdata
&& \multicolumn{2}{c}{AB} \\
\hline
Distance & pc & \multicolumn{2}{c}{$153\pm7^1$} \\
$J$ & mag & \multicolumn{2}{c}{11.86$^2$} \\
$H$ & mag & \multicolumn{2}{c}{11.14$^2$} \\
$K_s$ & mag & \multicolumn{2}{c}{10.76$^2$} \\
$\Delta J$ & mag & \multicolumn{2}{c}{$0.23\pm0.01$} \\
$\Delta K_p$ & mag & \multicolumn{2}{c}{$0.28\pm0.01$} \\
Angular separation, $\rho$ & arcsec & \multicolumn{2}{c}{$0.126\pm0.004$} \\
Position angle & degree & \multicolumn{2}{c}{$-80.99\pm0.10$} \\
\hline
&& Aab & Bab \\
\hline
Period, $P$ & days & 17.9420 $\pm$ 0.0012 & $4.541710\pm0.000019^3$ \\
Eccentricity, $e$ & \nodata & 0.2998 $\pm$ 0.0041 & 0.3224 $\pm$ 0.0042$^3$\\
Epoch, $T_o$ & BJD & 2457175.182 $\pm$ 0.031 & $2456896.19699\pm0.00019^3$\\
Longitude of periastron, $\omega$ & degree & 316.36 $\pm$ 0.93 & 100$^\ast$ \\
Semimajor axis, $a$ & AU & 0.09616 $\pm$ 0.00044 & 0.0359$^\ast$ \\
Minimum system mass, $(M_1+M_2)\sin^3i$ & \msun & 0.3685 $\pm$ 0.0050 & 0.3$^\ast$ \\
Mass ratio, $q$ & \nodata & 0.8309 $\pm$ 0.0062 & 1.0$^\ast$ \\
Obliquity, $\lambda$ & degree & \nodata & $-57^{+40}_{-36}$ \\

\enddata
\tablecomments{1,~\citet{GaiaDR2}, 2,~\citet{Cutri2003}, 3,~\citet{David2016}, $\ast$: guessed values used in obliquity measurement (\S \ref{sec:oblqt}).
}
\end{deluxetable*}

\begin{deluxetable*}{lcc}
\tablecaption{Parameters of the EPIC 203868608 Aab spectroscopic binary \label{table:epic608sb2}}
\tablecolumns{3}
\tablewidth{-0pt}
\tabletypesize{\scriptsize}
\tablehead{
        \colhead{Parameter} &
        \colhead{Units} &
        \colhead{Value} \\
        }
\startdata
\textit{Directly measured parameters} \\
Orbital period, $P$ & days & 17.9420 $\pm$ 0.0012 \\
Epoch, $T_0$ & BJD & 2457175.182 $\pm$ 0.031 \\
Primary Doppler semi-amplitude, $K_1$ & \kms &  26.46 $\pm$ 0.16  \\
Secondary Doppler semi-amplitude, $K_2$ & \kms & 31.84 $\pm$ 0.18 \\
Systemic radial velocity, $\gamma$ & \kms & -4.436 $\pm$ 0.072 \\
Eccentricity, $e$ &  & 0.2998 $\pm$ 0.0041 \\
Longitude of periastron, $\omega$ & deg &  316.36 $\pm$ 0.93 \\
RMS of primary RV fit & \kms & 0.6 \\
RMS of secondary RV fit & \kms & 0.8 \\
$\chi^2_\mathrm{red}$ of primary RV fit &  & 2.2 \\
$\chi^2_\mathrm{red}$ of secondary RV fit & & 2.7 \\
\\
\textit{Derived parameters} \\
Mass ratio, $q$ & & 0.8309 $\pm$ 0.0062 \\ 
Minimum system mass, $(M_1+M_2)\sin^3i$ & \msun & 0.3685 $\pm$ 0.0050 \\
Orbital separation, $a$ & AU & 0.09616 $\pm$ 0.00044 \\
\enddata
\end{deluxetable*}


\section{Summary and Discussion}
\label{sec:summary}

We provide observational evidence that EPIC 203868608 is a quadruple stellar system in the Upper Scorpius OB association. The system consists of two visual components, one being an SB2 and the other being an EB. All stellar components are consistent with being young low-mass stars (age $\sim$10 Myrs and individual masses lower than $\sim$0.3\msun) that are still undergoing pre-main sequence contraction.

Our observations include: Keck NIRC2 observations that spatially separates the two visual components and confirms the west component as the EB (\S \ref{sec:ao}); Keck HIRES observations that constrains the total mass and eccentricity for the SB2 and therefore the total mass of the system (\S \ref{sec:masses}); Keck NIRSPAO observations that indicates an inclined orbit for the EB (\S \ref{sec:oblqt}). The system represents a rare opportunity to test theories on star formation and subsequent dynamical evolution (\S \ref{sec:imp}). 

We would place a cautionary note here. The NIRSPAO observation is very challenging: namely to spatially separate a 0.126$^{\prime\prime}$ binary and measure LP changes for faint low mass stars. As a result, only 3 data points have been taken, out of which an inclined orbit of EPIC 203868608 B is inferred. Future AO-aided high-resolution spectroscopy observations are necessary to confirm this result. Available instruments include but are not limited to the upgraded NIRSPEC~\citep{Martin2014}, CRIRES+~\citep{Follert2014}, the Infrared Doppler instrument~\citep[IRD,][]{Kotani2014}, and the Keck Planet Imager and Characterizer~\citep[KPIC,][]{Mawet2016}. 

We develop a framework to infer orbital obliquity for eclipsing systems such as transiting planets and eclipsing stars. The framework includes an analytic approach to model LPs during an eclipse (\S \ref{sec:lpm}). Unlike previous analytic models, the analytic model can properly handle spectrally-resolved and/or self-luminous occulters, which is essential to modeling EBs and large planets around small stars. Additionally, the framework includes a matrix-based method to retrieve LPs and their uncertainties from high-resolution spectroscopic data (\S \ref{sec:lps}). Together, the framework offers an efficient way of inferring obliquity (\S \ref{sec:lpr}) and conducting simulations to check the robustness of the inference (\S \ref{app:sensitivity}). 

The framework will be particularly useful in studying the orbital architecture of eclipsing systems in the era of \textit{Kepler} and \textit{TESS} when numerous transiting planets and EB systems are being and will be discovered. 

\noindent{\it Acknowledgements} 
We thank the anonymous referee for his or her comments and suggestions that significantly improve the manuscript. We acknowledge Konstantin Batygin, Jim Fuller, and Fred Adams for discussions on tidal evolution and Kozai-Lidov perturbation in EPIC 203868608. We thank the PIs of KOA NIRC2 data, Christoph Baranec and Andrew Mann. SA acknowledges support by the Danish Council for Independent Research, through a DFF Sapere Aude Starting Grant No. 4181- 00487B. Funding for the Stellar Astrophysics Centre is provided by The Danish National Research Foundation (Grant agreement no.: DNRF106). Part of this research was carried out at the Jet Propulsion Laboratory, California Institute of Technology, under a contract with the National Aeronautics and Space Administration. TJD acknowledges support from the JPL Exoplanetary Science Initiative. The data presented herein were obtained at the W. M. Keck Observatory, which is operated as a scientific partnership among the California Institute of Technology, the University of California and the National Aeronautics and Space Administration. The Observatory was made possible by the generous financial support of the W. M. Keck Foundation. The authors wish to recognize and acknowledge the very significant cultural role and reverence that the summit of Maunakea has always had within the indigenous Hawaiian community.  We are most fortunate to have the opportunity to conduct observations from this mountain.

\bibliographystyle{apj}
\bibliography{mybib_JW_DF_PH5}

\appendix
\section{A: Sensitivity of NIRSPAO Observations to the RM Effect}
\label{app:sensitivity}
We investigate if we can robustly measure the obliquity and how well we can retrieve the information in the presence of noise. In order to answer the two questions, we run simulations to (1), generate LPs that are affected by the RM effect; (2), generate mocked observational data based on the LPs and with realistic noise; (3), use the same package as described in \S \ref{sec:oblqt} to analyze the mocked data.  

\begin{figure*}
  \centering
  \begin{tabular}[b]{cc}
    \includegraphics[width=.50\linewidth]{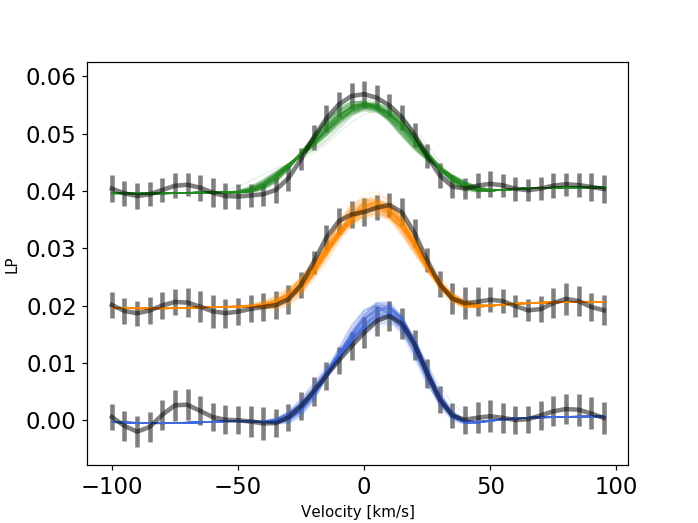} & \includegraphics[width=.45\linewidth]{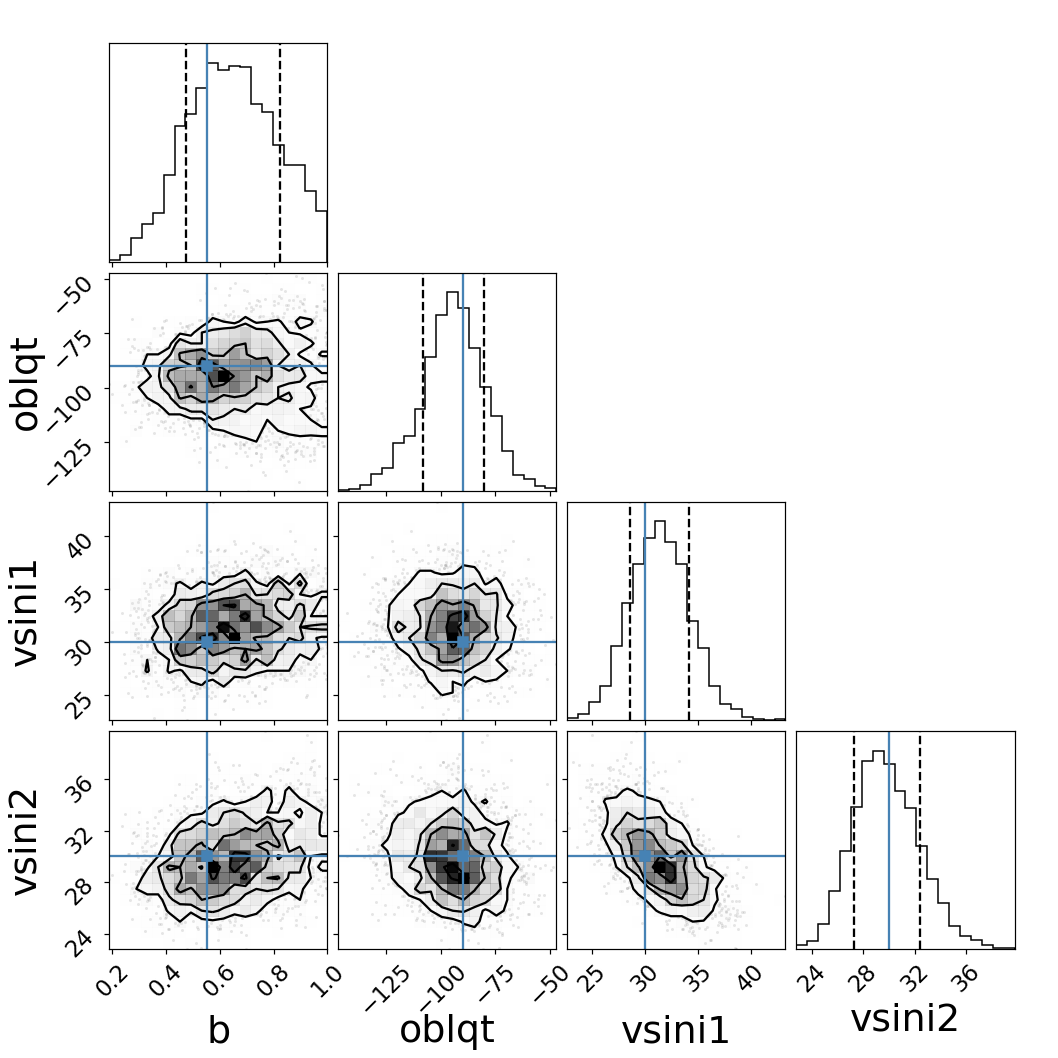} \\

  \end{tabular} \qquad
  \caption{Left: line profiles at different epochs for mocked data with an obliquity of -90$^\circ$. Grey data points with error bars are mocked data. Colored lines are random draws from posterior distributions. Right: posterior distribution of model parameters. Dashed lines mark 16 and 84 percentiles. Blue lines mark input values. 
  \label{fig:app_fig1}}
\end{figure*}

We consider a polar orbit with an obliquity of -90$^\circ$. We use the same model as described in \S \ref{sec:lpm} to model LPs. All the parameters in the model are the same as the measured ones except for the obliquity. Simulated LPs are shown on the left panel of Fig. \ref{fig:app_fig1}. 

Next, we simulate NIRSPAO observations as follows. We convolve a template spectrum (see \S \ref{sec:lpm}) with the mocked LP and multiply the convolution with telluric spectrum. The telluric spectrum is a convolution of a telluric template and a telluric LP that is measured simultaneously as the object LP~\citep{Wang2017b}. Finally, noise, which is the residual of modeling real data, is added to the mocked data. Using the residual as noise accounts for both photon noise and systematic noise such as detector noise and modeling error. 

We then infer the obliquity by analyzing the mocked data with the package as described in \S \ref{sec:oblqt}. Posterior distributions of model parameters are shown on the right panel of Fig. \ref{fig:app_fig1}. With an input obliquity of -90$^\circ$, the Bayesian inference returns an obliquity of $-94.1^{-14.0}_{+14.3}$ degree. Therefore, our inferred obliquity is consistent with the input value. 

The uncertainty of obliquity for the mocked data is a factor of $\sim$2-3 smaller than the reported value from the real data. We attribute the difference to a few steps in our analysis. First, the template spectrum we use does not perfectly agree with the real spectrum for a low-mass star, but it is assumed that they match in simulation. Second, our analytic model to describe an LP is imperfect as it parameterizes the Doppler shadow as the maximum and minimum velocities of the occulted area (see \S \ref{sec:lpm}), but reality may be more complicated. The mismatch is not accounted for in our simulation. Lastly, we fix the y-axis scaling factor in the analysis for mocked data. This is because LP normalization for mocked data has less uncertainty, which mainly comes from ripples outside the rotational broadening velocities. The real data show more ripples (Fig \ref{fig:LP}) than the mocked data (Fig. \ref{fig:app_fig1}), and therefore require an additional scaling factor to account for the normalization uncertainty.     


\section{B: Comparing Analytic Model with Finite-Element Model}
\label{app:comp}

We compare our analytic model to a finite-element model that pixelates stellar surface and numerically integrate to calculate LPs~\citep[][and references therein]{Albrecht2014}. Two cases are considered, one is a prograde orbit with $\lambda=0^\circ$ and the other case is a polar orbit $\lambda=-90^\circ$. All model parameters, except for obliquity, are the same as what are assumed or inferred from EPIC 203868608 Bab. Both models assume no macro-turbulence, but the finite-element model assumes a micro-turbulence of 2 \kms. Micro-turbulence and macro-turbulence do not play a significant role in the EPIC 203868608 Bab case because LP is mainly dominated by rotational broadening. 

\begin{figure*}
  \centering
  \begin{tabular}[b]{cc}
    \includegraphics[width=.48\linewidth]{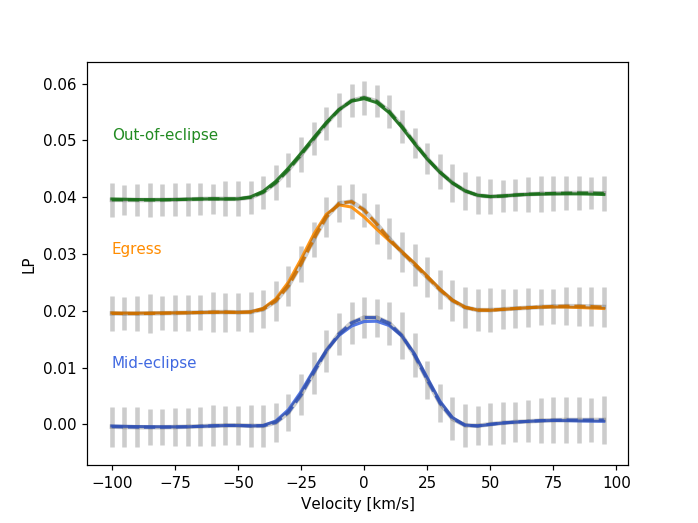} & \includegraphics[width=.48\linewidth]{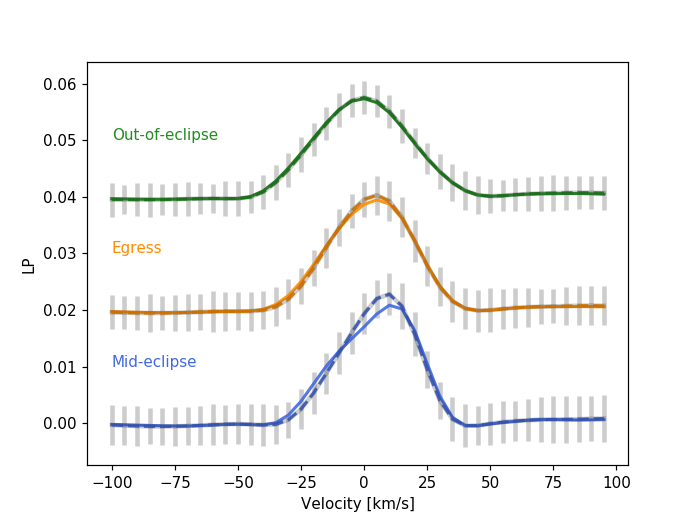} \\
  \end{tabular} \qquad
  \caption{Left: comparison between our analytic model (dashed lines) and a finite-element model~\citep[solid lines, ][and references therein]{Albrecht2014} for $\lambda=0^\circ$. LP measurement error bars are shown in light grey. Right: the same as left except for $\lambda=-90^\circ$. 
  \label{fig:app_fig2}}
\end{figure*}

Fig. \ref{fig:app_fig2} shows the comparison between the two models. In both the prograde and the polar orbits cases, the two models agree with each other within 7\% of their maximum values. The disagreement between models is at most 38\% of LP measurement errors (light grey bars in Fig. \ref{fig:app_fig2}). This suggests that LP measurement uncertainties dominate the uncertainty in our obliquity inference. However, the disagreement between models will be a source of systematic error for future observations with higher SNR and lower LP measurement uncertainty. The comparison between models shows that the analytic model used here needs to be improved. This issue will be addressed in a future paper~\citep[][\it{in prep}]{Pezzato2018}. 

\clearpage

\end{document}